\documentstyle[aasms4,11pt,epsf]{article}

\begin{document}

\title{The Unusual Near-Infrared Morphology of the Radio Loud Quasar 4C $+$09.17$^{*}$}
\author{L. Armus$^1$, G. Neugebauer$^1$, M.D. Lehnert$^2$, K. Matthews$^1$}
\affil{$^1$ Palomar Observatory, Caltech, Pasadena, CA 91125}
\affil{$^2$ Sterrewacht Leiden, Postbus 9513, 2300RA Leiden The Netherlands}
\affil{$^*$ Based on observations with the W.M. Keck Observatory, which is operated by the California Institute of Technology and the University of California}

\begin{abstract} 

Near-infrared images of the luminous, high redshift ($z=2.1108$) radio
loud quasar 4C $+$09.17 reveal a complex structure.  The quasar (K$=15.76$
mag) is surrounded by three ``companion" objects having 17.9 $<$ K $<$
20.2 mag at radii of 1.7$'' < \Delta$r$< 2.9''$, as well as bright,
diffuse emission.  The brightest companion has a redshift of $z=0.8384$
(Lehnert \& Becker 1997) and its optical-infrared colors (Lehnert et al.
1997) are consistent with a late-type spiral galaxy at this redshift with
a luminosity of about 2L$^{\ast}$.  This object is likely the galaxy
responsible for the strongest MgII absorption line system seen in the
spectrum of 4C $+$09.17 by Barthel et al.  (1990).  Redshifts are not
available for the remaining two companions.  The red colors of the second
brightest companion appear most consistent with a high redshift
star-forming galaxy at $z > 1.5$.  If this object is at the redshift of 4C
$+$09.17 it has a luminosity of about 7L$^{\ast}$.  The faintest companion
has colors which are unlike those expected from either a spiral or an E/S0
galaxy at any redshift associated with the 4C $+$09.17 system.  Since this
object lies along the same direction as the radio jet/lobe of 4C $+$09.17,
as well as the extended Ly$\alpha$ emission mapped by Heckman et al.
(1991) we suggest that this component can be explained as a combination of
strong line emission and scattered QSO light.  The resolved, diffuse
emission surrounding 4C $+$09.17 is bright, K$\sim17.0$ mag, and about one
magnitude redder in J-K than the quasar.  If this emission is starlight, a
very luminous elliptical host galaxy is implied for 4C $+$09.17.
Scattered and reddened AGN light, emission line gas, and flux from
absorbing galaxies along the line of sight may all contribute to this
emission.

\end{abstract}


\section{Introduction}\label{sec:introduction}

At low redshifts, quasars often sit at the centers of luminous galaxies
(Hutchings, Crampton \& Campbell 1984, Smith et al. 1986, Bahcall et al.
1997).  In general, radio loud quasars, those with monochromatic radio
powers at centimeter wavelengths of more than $10^{25}-10^{26}$ W
Hz$^{-1}$ and radio-to-optical monochromatic luminosity ratios larger than
$\sim100$ - see Kellerman et al. (1988) and Schneider et al. (1992),
inhabit elliptical galaxies, while radio quiet quasars tend to be found in
spiral galaxies (Boroson \& Oke 1984; Malkan 1984; Boroson, Persson \& Oke
1985; Smith et al.).  Furthermore, the host galaxies of radio loud quasars
tend to be more luminous by about one magnitude in the visual than their
radio quiet counterparts (Smith et al.).  The reason for these apparent
dichotomies are not well understood.  Images of low redshift luminous
(M$_{B} < -23$ mag) quasars obtained with the Hubble Space Telescope show
a wide variety of host galaxy morphologies including both elliptical and
spiral systems as well as highly disturbed galaxies which may be
interacting with and/or accreting close companions (Bahcall et al., Disney
et al. 1995, Hutchings \& Morris 1995).  In addition, recent deep infrared
imaging (Neugebauer, Matthews \& Armus 1995 ; Mcleod \& Rieke 1995) has
proven very effective in finding smooth host galaxies and faint companions
around some low redshift quasars.  That the nature of quasar host galaxies
even at low redshift is still a matter of some debate is evidenced by the
recent discussions concerning the Hubble Space Telescope results presented
by Bahcall et al., and the efforts made to reconcile these results with
ground based measurements made at longer wavelengths (e.g. McCleod \&
Rieke).  In particular, the fact that 50\% of the radio quiet quasars
imaged by Bahcall et al. seem to have elliptical hosts has challenged the
apparent clear-cut distinction between the hosts of radio quiet and radio
loud quasars.

At redshifts above $z\sim2$, dis-entangling host galaxy light from AGN
light is made difficult by the relative faintness of the host galaxy as
well as the blurring effects of atmospheric seeing.  However, it is
precisely at these redshifts that studies of the near environments of
quasars offer the most hope of furnishing clues to the early evolution of
AGN and galaxies.  It is now well established that the redshift regime of
$z\sim2-3$ represents a real maximum in the number density of luminous
quasars (Hartwick \& Schade 1990; Schmidt, Schneider \& Gunn 1995).
Between these redshifts and the present, the number density of luminous
quasars has dropped by more than a factor of 1000, suggesting very strong
cosmic evolution.  If high redshift quasars sit at the centers of luminous
host galaxies, the peak in the quasar density at $z\sim2-3$ may signal an
epoch where luminous galaxies were being hierarchically assembled (e.g.
Carlberg \& Couchman 1989).

As part of a K-band imaging survey of the Mpc-scale environments of a
large sample of high redshift quasars with the W.M. Keck Telescope, we
have imaged a very complex system around the $z=2.1108$, radio loud
quasar, 4C $+$09.17 at RA(1950)$=$04h45m37.12s,
DEC(1950)$=+09$d45'37.2$''$.  This object serves as an excellent example
not only of the wealth of detail that a deep imaging program of this
magnitude will provide, but also of the various pitfalls associated with
inferring the properties of the host galaxies and companions of high
redshift quasars.

4C $+$09.17 was previously found to have an extended, asymmetric
Ly$\alpha$ nebula with a luminosity of about $1.6\times10^{44}$ erg
s$^{-1}$ by Heckman et al. (1991).  Lehnert et al. (1992) imaged 4C
$+$09.17 in the K-band and found resolved emission at radii from 1$''$ to
6$''$ with a K$\sim 17.5\pm0.4$ mag.  This emission appeared asymmetric,
but the true morphology of the nebulosity was unclear.  The Keck J and
K-band data presented here show a group of three objects within 3$''$ of
the quasar as well as very bright diffuse near-infrared emission and they
allow a determination of the magnitudes and colors for each of the
components.

Barthel et al. (1990) quote a number of emission line redshifts for 4C
$+$09.17 from the CIV, HeII and CIII] features, spanning a range of
$z=2.1085$ to $z=2.1114$.  We adopt the average emission line redshift of
$z=2.1108$, as determined by Barthel et al.  as the systemic redshift of
4C $+$09.17.  In addition to the emission lines, there are four absorption
line systems in the spectrum of 4C $+$09.17 (Barthel et al.).  Two of
these are MgII systems at redshifts of $z=0.839$ and $z=1.466$, the former
being significantly stronger than the latter, with rest frame equivalent
widths (EQW) of $2.68\AA$ vs.  $0.46\AA$, respectively.  In addition,
there is a CIV absorption system at a redshift of $z=2.107$, having a rest
frame EQW of $2.3\AA$.  Finally, there is a CIV absorption line system
with $z_{abs} \sim z_{em}$ and a rest frame EQW of $1.4\AA$.

Throughout this paper we adopt H$_{o}=75$ km s$^{-1}$ Mpc$^{-1}$ and
q$_{o}=0$, so that the distance to 4C $+$09.17 is 5576 Mpc, and 1$''$ on
the sky is equivalent to 8.8 kpc in projection.  In order to compare the
observed infrared magnitudes to the expected rest frame values of typical
galaxies in the local Universe, we employ the optical luminosity function
of Mobasher, Sharples \& Ellis (1993), such that a spiral galaxy has a
characteristic absolute magnitude, M$_{B}^\ast = -20.49$ mag, while an
E/S0 galaxy has a characteristic absolute magnitude, M$_{B}^\ast = -20.24$
mag, following Schecter (1976), and converting the Mobasher, Sharples \&
Ellis values into ones appropriate for an H$_{o}=75$ km s$^{-1}$
Mpc$^{-1}$.  Combining these characteristic absolute magnitudes with the
spectral energy distributions of Coleman, Wu \& Weedman (1980), we
estimate that the apparent K magnitudes of an L$^{\ast}$ spiral and
E/S0 galaxy at a redshift of $z=2.1108$ are K$=21.46$ mag and K$=20.99$
mag, respectively.

\section{Observations and Data Reduction}\label{sec:observations} 

Observations of 4C $+$09.17 were made at the W.M. Keck Observatory on
the nights of 13 Feb 1995 and 5 Oct 1995 UT with the Near Infrared
Camera (NIRC).  The quasar was imaged through a broad-band K filter
(2.0-2.45$\mu$m) during both runs, and through a broad-band J filter
(1.12-1.37$\mu$m) during the Oct 1995 observing run.  The plate scale
of the 256x256 InSb array is 0.15$''$ per pixel.  The total
integration time through the K filter was 1080 seconds in Feb 1995
and 540 seconds in Oct 1995.  The total integration time through the
J filter was 540 seconds in Oct 1995.  In each case, individual
images of 60 seconds duration have been taken with the quasar moved
by $\sim10''$ on the array between successive exposures.  Sky and
flat field frames were generated from the data by combining exposures
in groups of $7-9$ using either a clipped mean or median filtering
technique.  An offset guider employing a visual wavelength CCD was
used to accurately maintain the telescope tracking.  Since the quasar
is visible in each 60 second exposure, centroids were measured for
the individual frames and used to derive the offsets used to create
the final K and J-band mosaics.  In addition to the quasar,
observations were made of a nearby ($\Delta r =11 \& 5$ arcminutes)
star either immediately before (Oct 95 data) or immediately after
(Feb 95 data) the observations of the quasar, respectively, in order
to obtain an estimate of the point spread function (PSF).  During
both observing runs the PSF star was imaged in a manner identical to
that used for the quasar itself, using the same raster pattern and
exposure times (60 seconds per position).  From 
these PSF images the seeing was estimated to be
approximately $0.7''$  and $0.5''$ full width at half maximum (FWHM)
during the Feb 1995 and Oct 1995 observations of 4C $+$09.17,
respectively.  The conditions were photometric during both observing
runs, and observations of UKIRT faint standard stars (Casali \&
Hawarden 1992) provided the flux calibration.  Although the seeing
was steady during the Oct 95 observing run, it degraded noticeably to
$0.9''-1.0''$ FWHM during the observations of the standard stars on
the Feb 1995 run, making small aperture fluxes derived from the
latter data set uncertain.  Although both data sets are combined to
obtain the deepest possible images of 4C $+$09.17 (Fig. 1), only the
Oct 95 data set is used to derive magnitudes for the individual
components and assess the nature of the extended emission in the 4C
$+$09.17 system (Figs. 2 \& 3).

Visual images of the 4C $+$09.17 system were obtained with the Hubble
Space Telescope using the Wide Field and Planetary Camera (WFPC2) on
9 September 1994. These images were taken through the the F555W
filter.  The observations of 4C $+$09.17, along with the results from
observations of five other high redshift QSOs with HST, are fully
described in Lehnert et al.  (1997).  The 4C $+$09.17 data are
included here in order to derive the visual-infrared colors of the
system.

The Galactic color excess towards 4C $+$09.17 is E(B-V)$\sim 0.16$ mag
(Burstein \& Heiles 1982). Using the reddening curves of Savage \& Mathis
(1979) and Rieke \& Lebofsky (1985), we derive a Galactic extinction of
A$_{V}=$0.5 mag, A$_{J}=$0.14 mag and A$_{K}=$0.06 mag, respectively.  The
data presented in the following sections are corrected for these
extinctions.

\section{Results}\label{sec:results} 

The infrared observations reveal a complex structure surrounding the
z=2.1108, radio loud quasar 4C $+$09.17.  The quasar resides in a system
containing three distinct components visible at 2.2 microns, in addition
to a resolved, diffuse nebula.  The composite K-band mosaic made from the
combined Feb 1995 and Oct 1995 data sets is shown in Fig. 1.  The
brightest companion, 1.72$''$ southeast of the quasar and hereafter
labelled as component ``B", is clearly resolved, with a structure that
appears elongated in the northeast-southwest direction.  This object has a
K magnitude of $17.87\pm 0.03$ mag, and a color of J-K $=1.56\pm 0.04$ mag
as measured through a 2.0$''$ diameter circular beam centered on the
emission peak.  For comparison, the quasar itself (component ``A") has a K
magnitude of $15.76\pm 0.03$ mag and a color of J-K $= 1.46\pm 0.04$ mag.
Component ``C", located 2.54$''$ northeast of the quasar, has a K
magnitude of $19.42\pm0.07$ mag, and a color of J-K $=2.16\pm0.15$ mag.
Component C is the reddest member of the 4C $+$09.17 system.  Finally,
there is a very faint, diffuse emission region approximately 3$''$
southwest of the quasar.  This emission has a local maximum, labelled ``D"
in Fig. 1, approximately 2.9$''$ from the quasar, with a K magnitude
$20.15\pm0.13$ mag and a J-K color of $1.17\pm0.17$ mag.  Component D is
found along the direction of the radio jet mapped at 15 GHz by Barthel et
al.  (1988) and Lonsdale, Barthel, \& Miley (1993), but is at a larger
distance from the quasar than the radio lobe at a radius of 1.6$'' -$
marked with a cross in the lower panel of Fig. 1.  The magnitudes and
colors of the quasar and its companions, corrected for Galactic
extinction, are listed in Table 1.

It is clear from Fig. 1 that the entire 4C $+$09.17 system is surrounded
by diffuse infrared emission.  In Fig. 2a, we plot the azimuthally
averaged K-band surface brightness of the Oct 1995 data out to a radius of
4.5$''$.  For comparison, the same quantity as a function of radius is
plotted for the PSF star taken immediately before the quasar
observations.  Both profiles have been normalized to the flux within the
central three pixels, or $0.45''$.  Excess emission above that expected
from a point source is clearly seen at all radii.  The ``bump" in the
quasar surface brightness distribution at about 1.5$''$ is caused by
component B.  For comparison, the K-band surface brightness of the quasar
and PSF star calculated over a restricted sky position angle of
$262^{\circ}<$ PA $<352^{\circ}$ (measured east from north) to exclude
components B, C and D from the average are plotted in Fig. 2b.  The excess
over the point spread function is still obvious when the contributions
from the quasar companions are removed.

{}From estimates of the the amount of resolved emission in the J and K-band
images, constraints can be placed on the luminosity and color of the
emission around 4C $+$09.17.  The FWHM of the PSF images were about
$0.5''$ in both the J and K filters during the Oct 95 observations.  An
estimate of the fraction of the total emission which is resolved around 4C
$+$09.17 has been made by scaling the flux found within a radius of
0.37$''$ (2.5 pixels) in the QSO image to that found in the PSF image.
Since there is asymmetric structure around 4C $+$09.17 due to components
B, C, and D, we have performed this calculation over the same restricted
angle used to generate Fig.  2b.  In the K-band, approximately $27\%$ of
the total emission beyond a radius of 0.37$''$ is resolved, while in the
J-band the fraction is about $11\%$.  The total magnitudes of the 4C
$+$09.17 system measured in 8$''$ diameter circular beams centered on the
QSO (corrected for Galactic extinction) are K$=15.45$ mag and J$=16.94$
mag.  After removing the contributions of components B, C, and D to the
total flux, the resolved emission around 4C $+$09.17 (beyond a radius of
0.37$''$) has K$\sim17.0$ mag and J$\sim19.5$ mag.  This is slightly
brighter than the measurement of Lehnert et al. (1992).

In Figure 3 the infrared and visual flux densities, normalized to the flux
density at an observed wavelength of $2.2\mu$m, are shown for components
B, C, and D, as well as the quasar itself.  These are overlayed on four
representative spectral energy distributions (an observed Im, Scd, Sbc,
and an E/S0 galaxy spectrum) taken from Coleman, Wu \& Weedman (1980)
which have been redshifted to $z=2.1108$ (Fig. 3a), and $z=0.84$ (Fig.
3b).  The rest-frame UV data have been corrected for Galactic extinction
by Coleman, Wu \& Weedman.

\section{Discussion}\label{sec:discussion}

The apparent quasar companions are most likely to originate in one of the
following ways.  First, each could be a foreground galaxy, possibly
associated with one or more of the absorption systems seen in the spectrum
of 4C $+$09.17 by Barthel et al. (1990).  The three $z_{abs} << z_{em}$
(MgII and CIV) absorption line systems in the quasar spectrum provide
potential redshifts for the nearby companions.  Second, any or all of the
companions could be at the same redshift as the quasar, possibly
interacting with or accreting onto the quasar host galaxy.  If they are at
the redshift of 4C $+$09.17, starlight, ionized gas, and/or scattered AGN
light could contribute to the measured infrared fluxes.  We deal with each
of the infrared components separately below in light of these
possibilities.

\subsection{Component B}

Component B, has recently been shown to be at a redshift of $z=0.8384$
from an identification of the [OII] 3727$\AA$ emission line by Lehnert \&
Becker (1997).  This redshift is consistent with the idea that component B
is responsible for the strongest MgII absorption line system in the
spectrum of 4C $+$09.17.  At this redshift, the separation on the sky of
component B and 4C $+$09.17 (the MgII ``impact parameter") is about 11.6
kpc.  The empirical relationship between impact parameter and rest frame
MgII equivalent width found by Lanzetta \& Bowen (1990), predicts a
separation of $8-10$ kpc between 4C $+$09.17 and the absorbing galaxy
producing the MgII absorption line, when their data are converted to an
H$_{\circ}=75$ km s$^{-1}$ Mpc$^{-1}$.  The optical-infrared colors of
component B are most similar to those of a late-type spiral galaxy at
$z\sim0.8$ (Fig. 3b).  In Fig. 4 the V-J vs.  J-K colors of the galaxy
templates used in Fig. 3 are plotted as a function of redshift from
$z=0-2.2$ in steps of $\Delta z=0.2$.  Also shown in Fig. 4 are three
reddening vectors corresponding to the change in J-K and V-J for a rest
frame A$_{V}=1.0$ mag at $z=0$, $z=1.0$ and $z=2.0$.  The galaxies used as
templates for the Im, Scd, Sbc and E/S0 Hubble types by Coleman, Wu \&
Weedman (1980) are shown as observed, i.e. they include intrinsic
reddening.  Component B can be reconciled with the Scd curve with a slight
amount (A$_{V}\sim0.5$ mag) of extra reddening.  The J-band magnitude of
component B, J$\sim19.4$ mag, corresponds to a monochromatic luminosity of
about 2L$^{\ast}$ at a rest frame wavelength of 6800$\AA$.  The infrared
and visual fluxes are therefore consisent with component B being a
late-type spiral galaxy at $z=0.84$.

\subsection{Component C}

Due to its proximity (at least in projection), it is natural to ask
whether component C could be associated with component B.  Component C
could be a galaxy at $z\sim0.8$, yet it is redder than even an old stellar
population at this redshift (see Figs. 3b \& 4).  The J-K and V-J colors
of component C imply that the source lies at a fairly large redshift,
i.e.  $z\ge 1.5$.  Although the (3$\sigma$) limit on the V magnitude
prohibits an accurate estimate of the V-J color of component C, it appears
that the colors can be matched by either the Im or the spiral galaxy
templates by varying the amount of reddening as long as the galaxy is at a
high redshft.  If component C is at the redshift of 4C091.7, it may be a
7L$^{\ast}$ spiral at a projected separation of about 18 kpc.  The colors
of component C are consistent however, given the V-J limit and the
uncertain reddening, with a spiral galaxy at the redshift of the
$z=1.4664$ MgII absorption line system in the spectrum of 4C $+$09.17.
Using the relationship between impact parameter and MgII absorption
equivalent width (Lanzetta \& Bowen 1990), the $0.46\AA$ equivalent width
would predict a separation of about 40 kpc between component C and the
quasar.  The measured separation is 2.1$''$, or only 17 kpc at $z=1.466$,
implying an expected rest frame equivalent width of $1.0-1.5\AA$.  If
component C is a spiral galaxy at $z=1.466$, it has a relatively weak MgII
absorption line system (for its measured impact parameter) and a
luminosity of about 2L$^{\ast}$.

\subsection{Component D}

Component D is fainter and bluer than components B and C.  The combination
of blue J-K and red V-K colors are quite unlike those expected for a
galaxy at the redshift of component B (see Figs. 3b and 4).  It is clear
from Fig. 4 that simply reddening any of the galaxy templates will not
reproduce the colors of D.  Within the uncertainties, the V-J colors of D
and B are the same, but D is about 0.3-0.4 mag bluer in J-K.  If component
D is at the redshift of component B, the H$\alpha$ line is redshifted to
1.208$\mu$m, and lies in the J-band atmospheric window.  This line (or
more specifically, the H$\alpha+$[NII] blend) could therefore contribute
to the blue J-K color of component D, since there is no correspondingly
strong line which is redshifted into the K-band window.  To decrease the
brightness of component D by 0.3 mag would require an H$\alpha+$[NII]
line flux of about 5.3$\times 10^{-16}$ erg cm$^{-2}$ s$^{-1}$,
corresponding to a luminosity of about 1.4$\times 10^{42}$ erg s$^{-1}$ at
$z=0.84$.  This hypothesized line would have an intrinsic equivalent width
of about 420$\AA$, large even for luminous, high redshift radio galaxies
(e.g. Eales \& Rawlings 1993).  Furthermore, if component D were a spiral
galaxy at $z=0.84$ it would be intrinsically very faint at about
0.3L$^{\ast}$.  The large H$\alpha+$[NII] luminosity required to account
for the blue J-K colors of D coupled with its intrinsically faint overall
luminosity make it unlikely that D is a galaxy at $z=0.84$.

Since D is located along the radio jet/lobe axis of 4C $+$09.17, and there
is known extended L$\alpha$ emission line gas at the location of component
D which is probably ionized by the quasar, it is unlikely that D is at
$z=0.84$. It is most plausible to suggest that component D is at the
redshift of 4C $+$09.17.  If D is at $z=2.1108$, H$\alpha$ can contribute
to the K band flux at $2.04\mu$m, and [OII] 3727$\AA$ can contribute to
the J band flux at $1.16\mu$m.  Heckman et al. (1991) find extended UV
continuum and Ly$\alpha$ emission surrounding 4C $+$09.17.  The resolved
Ly$\alpha$ flux is $\sim4\times10^{-15}$ erg cm$^{-2}$ s$^{-1}$, corrected
for a Galactic exctinction of 0.7 mag at $3800\AA$ (Burstein \& Heiles
1982, Savage \& Mathis 1979).  If the ionized gas around 4C $+$09.17 has
the same Ly$\alpha$-to-H$\alpha$ emission line flux ratio as found in high
redshift radio galaxies, i.e. 3.5 (McCarthy, Elston \& Eisenhardt 1992),
the expected resolved H$\alpha$ flux is approximately $10^{-15}$ erg
cm$^{-2}$ s$^{-1}$.  Heckman et al. note that the resolved Ly$\alpha$
emission in 4C $+$09.17 is not distributed symmetrically, but is extended
more towards the southwest - the direction of both the radio jet and
component D.  Inspection of their Fig. 1 shows that the radio lobe and
component D lie well within the brightest extended Ly$\alpha$ emission
line nebula.  It is likely therefore, that any emission line contribution
to component D comes from gas ionized by the quasar.

As an upper limit to the contribution of emission lines to the K-band flux
of component D, we can assign to it all the resolved H$\alpha$ emission
line flux estimated from the measured, resolved Ly$\alpha$ emission.  When
this is done, nearly 73\% of the flux from component D could be from
H$\alpha$, and the corrected continuum light has a K$\sim20.8$ mag.
Performing the same calculation for [OII], and assuming the
Ly$\alpha$-to-[OII] emission line flux ratio is the same in 4C $+$09.17 as
it is in radio galaxies, namely about 8.5 (McCarthy 1993), we estimate
that at most $22\%$ of the J-band light seen from component D could be
[OII] emission - not as large a fraction as estimated for H$\alpha$, but
still significant.

Subtracting the maximum emission line contributions from the measured J
and K-band fluxes makes the continuum J-K color of component D bluer than
the redshifted Im galaxy spectral energy distribution in Fig. 3a.  If the
maximum emission line contribution is used, the continuum from component D
is too blue to be starlight at the redshift of 4C $+$09.17.  This
conclusion is strengthened if resonant scattering of Ly$\alpha$ photons is
an important effect, since then the H$\alpha$ line flux would be larger
than we have estimated by using the measured Ly$\alpha$ line flux from
Heckman et al. (1991).  On the other hand, the H$\alpha$ emission line
contribution to the K-band flux of component D may be an upper limit for
two reasons.  First, the Ly$\alpha$ nebula mapped by Heckman et al.
appears larger and more diffuse than component D, so the line emission is
probably not concentrated in one area.  Second, the H$\alpha$ emission
line is near the edge of the K-band atmospheric window at $2.04\mu$m, and
thus the contribution of this line to the integrated K magnitude is
dependent upon the kinematics of the emission line gas.  However, it
nonetheless appears that if component D is at the redshift of 4C $+$09.17
and emission lines are contributing to the broad band near infrared
fluxes, the blue continuum colors are inconsistent with starlight.  If
component D is at the redshift of 4C $+$09.17, its continuum flux may be
dominated by reflected AGN light and emission lines energized by the
quasar.

\subsection{The Diffuse Emission around 4C $+$09.17}

In most cases, symmetric, diffuse emission around high redshift quasars is
assumed to be at the redshift of the quasar since the likelihood of
finding a foreground faint galaxy perfectly centered on the background
quasar is very small.  The symmetric emission can be starlight, emission
line gas, or scattered AGN light.  The complex morphology of 4C $+$09.17
and the coincidence of strong, redshifted emission lines forces us to
consider all of these possibilities.

The resolved, diffuse emission beyond a radius of 0.37$''$ (Fig. 2b) has a
magnitude of K$\sim17.0$ mag and J-K$\sim2.5$ mag.  This is redder in J-K
than any other component in the 4C $+$09.17 system.  The red J-K color of
the extended emission seems inconsistent with a strong contribution from
scattered AGN light at these wavelengths, unless the scattered light is
heavily reddened by dust along the line of sight.  If the diffuse emission
is starlight, it is extremely luminous, and most similar in color to an
evolved stellar population.  A K-band magnitude of K$\sim17.0$ mag implies
a 39L$^{\ast}$ E/S0 host galaxy for 4C $+$09.17.  This luminosity is in
fact a lower limit on the true luminosity of the 4C $+$09.17 host galaxy,
since the measurement is made beyond a radius of 0.37$''$ from the
quasar.  At $z=2.1108$, 0.37$''$ is about 3.2 kpc.  If the host galaxy
follows an r$^{1/4}$ light profile (de Vaucouleurs 1948) with an effective
radius of about 4 kpc (Kormendy 1977), the total luminosity of the 4C
$+$09.17 host galaxy could be as much as $70-80$L$^{\ast}$.  Although
extended emission around other high redshift quasars has been taken as
evidence for extremely luminous ($>20$L$^{\ast}$) host galaxies by Lehnert
et al.  (1992), in the case of 4C $+$09.17 there are known absorption line
systems along the line of sight, implying that some of the light around
the quasar must be in the foreground.  Component B, which we have
associated with the MgII absorption system at $z=0.84$ is clearly
extended.  If this galaxy is symmetric, at least in the direction toward
and away from 4C $+$09.17, we estimate that about 10\% of its total
emission lies at a radius of greater than 2.0$''$, i.e. at a large enough
distance to contribute to the diffuse emission around the quasar.  This
flux from component B could be responsible for about 6\% of the resolved
emission around 4C $+$09.17.  If component B is highly asymmetric toward
4C $+$09.17, perhaps a result of a galactic arm or tidal tail, the
contribution could be larger than 6\%, although it seems not to be a
dominant effect.

There is certainly resolved emission line gas around 4C $+$09.17, as
evidenced by the spatially resolved Ly$\alpha$ imaging presented by
Heckman et al.  (1991).  In principle, H$\alpha$, [NII], and [SII]
emission may contribute to the integrated K-band light seen surrounding
the quasar.  However, even if all the resolved Ly$\alpha$ emission seen by
Heckman et al.  is used to estimate the resolved H$\alpha$ line flux, less
than 5\% of the resolved K-band emission around 4C $+$09.17 could be due
to this emission.

\section{Summary}\label{sec:summary}

Near-infrared images of the radio loud quasar 4C $+$09.17 show the
presence of three ``companion" objects with brightnesses of 17.9 $<$ K $<$
20.2 mag, as well as resolved, diffuse emission surrounding the quasar.
The brightest of these components has been measured to have a $z=0.8384$
(Lehnert \& Becker 1997) and is probably the galaxy responsible for the
strong MgII absorption seen in the spectrum of the quasar.  The
optical-infrared colors of this object are consistent with a late-type
spiral galaxy at this redshift, with a luminosity of about 2L$^{\ast}$.
The second brightest companion at K appears most consistent with a
star-forming galaxy at $z > 1.5$, and if it is at the redshift of 4C
$+$09.17, it has an intrinsic luminosity of about 7L$^{\ast}$. If on the
other hand, it is a spiral galaxy at a $z\sim1.5$ it may be associated
with the weaker MgII absorption line system at $z=1.466$, and have a
luminosity of about 2L$^{\ast}$.  The faintest member of the 4C $+$09.17
system has very blue J-K, yet red V-J colors, which appear inconsistent
with either an old or young stellar population at the absorption or
emission line redshifts associated with 4C $+$09.17.  The unusual colors
and diffuse morphology of this component may be explained if it is a
combination of strong line emission and scattered QSO light.  The fact
that it lies along the direction of the radio jet and the extended
Ly$\alpha$ emission line gas strengthens this interpretation.  The diffuse
emission surrounding 4C $+$09.17 is very bright (K$\sim17.0$ mag) and
redder than the QSO by about one magnitude in J-K.  If this emission is
starlight at the redshift of 4C $+$09.17, an elliptical host galaxy with a
luminosity larger than about 40L$^{\ast}$ is implied.  Scattered and
reddened AGN light, emission line gas, and flux from absorbing galaxies
along the line of sight may all contribute to this emission.

\acknowledgments

The W.M. Keck Observatory is operated as a scientific partnership between
the California Institute of Technology and the University of California.
We thank the entire Keck staff, especially Wendy Harrison and Al Conrad,
for making these observations possible.  In addition, we thank Peter
Barthel, Tim Heckman, David Hogg, Neill Reid, Tom Soifer, and Chuck
Steidel for helpful discussions.  Finally, we would like to thank Steve
Warren, the referee, whose comments and suggestions helped to improve the
final version of this paper.  Infrared astrophysics at Caltech is
supported by grants from NASA.  This research has made use of the
NASA/IPAC Extragalactic Database which is operated by the Jet Propulsion
Laboratory, Caltech, under contract with NASA.

\clearpage

\begin{table}
\center{\bf Components in the 4C $+$09.17 System}
\begin{center}
\begin{tabular}{ccccc}
Component&$\Delta$r&K&J-K&V-K\cr
 &$''$&mag&mag&mag \cr
\hline
\hline
A$=$QSO&0.0&15.76$\pm0.03$&1.46$\pm0.04$&3.57$\pm0.04$ \cr
B      &1.7&17.87$\pm0.03$&1.56$\pm0.04$&4.69$\pm0.07$ \cr
C      &2.1&19.42$\pm0.07$&2.16$\pm0.15$&$>5.35$ \cr
D      &2.9&20.15$\pm0.13$&1.17$\pm0.17$&4.40$\pm0.24$ \cr

\end{tabular}
\caption{Near infrared magnitudes and colors of the components in the 
4C $+$09.17 system.  All values are measured through circular beams of 
2.0$''$ diameter.
Column 1 is the designation used in the text.
Column 2, $\Delta r$, is the distance of the component from the quasar.}
\end{center}
\end{table}

\clearpage

\begin{center}
{\bf Figure Captions}
\end{center}

\figcaption[0445_plot1.ps]{Two views of 4C $+$09.17 in the K-band.  In the
top panel, the color palette has been chosen to highlight the quasar
(component ``A") and the brightest of the near-infrared companions,
component ``B".  A solid bar depicting a projected scale of 20 kpc at the
redshift of the quasar (for H$_{\circ}=$75 km s$^{-1}$ Mpc$^{-1}$ and
q$_{\circ}=0$) is also shown.  In the bottom panel, the color palette has
been chosen to highlight the diffuse emission surrounding the system, as
well as components ``C" (at 1,2) and ``D" (at -2,-2).  A cross marks the
position of the radio lobe mapped by Barthel et al. (1988), and Lonsdale,
Barthel \& Miley (1993), assuming that the brightest radio component is
coincident with the quasar.  Component ``D" lies along the same direction
from the quasar as the radio jet, yet at a larger distance than the radio
lobe.  In both images, north is up and east is to the left.}

\figcaption[prof2.ps]{Azimuthally averaged K-band surface brightness
profile of 4C $+$09.17 in units of relative flux per pixel.  In (a) the
light is averaged over all angles.  In (b) the average is restricted to
$262^{\circ}<$PA$<352^{\circ}$, chosen to exclude the contribution from
components B, C, and D from the average.  In both cases the quasar is
shown as solid squares and the point spread function (PSF) is shown as
open squares.  The quasar and the PSF have been normalized to contain the
same number of counts within the central $0.45''$.}


\figcaption[plot3.ps]{The optical-near infrared flux densities (normalized
to the K-band flux density) of the components in the 4C $+$09.17 system,
corrected for a Galactic extinction of A$_{V}=0.5$ mag, and overplotted on
representative galaxy spectral energy distributions taken from Coleman, Wu
\& Weedman (1980).  The Im galaxy spectrum is a composite of NGC 4449 and
NGC 1140, the Scd galaxy spectrum is a composite of M33 and NGC 2403, the
Sbc galaxy spectrum is a composite of M51 and NGC 2903, and the E/S0
galaxy spectrum is a composite of the bulges of M31 and M81.  All UV
spectral points have been corrected for Galactic extinction.  The energy
distributions have been extended into the near infrared for fig. 3b by
matching the J-H and H-K colors of spiral and elliptical galaxies from
Aaronson (1977) to the I-J colors of models with the closest SED match in
Bruzual \& Charlot (1993).  Two redshifts are depicted here, $z=2.1108$
(Fig 3a), and $z=0.839$ (Fig 3b), corresponding to the redshift of the
qusar as well as the strongest MgII absorption line system seen in the
spectrum of 4C $+$09.17 by Barthel, et al.  (1990), respectively.  The
observed F555W point for component C is an upper limit.  The diamond
symbol in each plot marks the center of the observed K-band window, to
which the individual spectra have been normalized.}

\figcaption[plot4.ps] {The V-J vs. J-K colors of components B, C, and D
overplotted on the colors expected from the basic galaxy types of Fig. 3
(from Coleman, Wu \& Weedman 1980) as a function of redshift between
$0<z<2.2$, in steps of $\Delta z=0.2$.  The Im galaxy is given by open
squares, the Scd galaxy by open triangles, the Sbc galaxy by open circles,
and the E/S0 galaxy by solid circles.  The data for components B, C, and D
are indicated by solid squares and labelled accordingly.  The Scd, Sbc,
and E/S0 points corresponding to $z=1$ and $z=2$ are labelled.  In
addition, three reddening vectors are shown corresponding to additional
extinctions of A$_{V}=1$ mag in the rest frame of an object at $z=0$,
$z=1$, and $z=2$.  The red colors of C indicate a high redshift galaxy,
while the colors of D are difficult to fit with any of the generic galaxy
types.  The optical-infrared colors of component B are consistent with a
late type spiral galaxy at a $z\sim0.84$.}

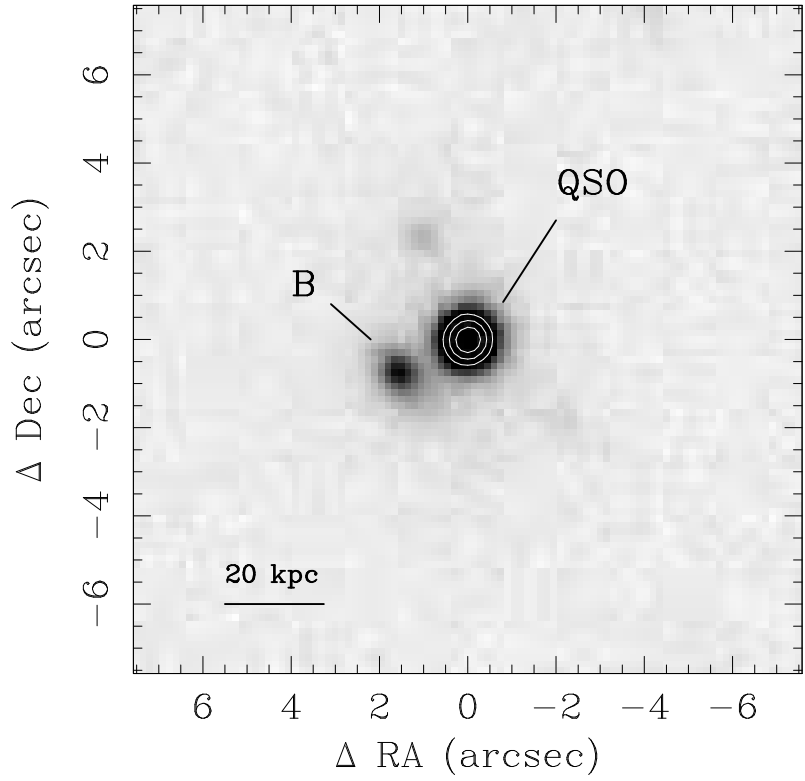
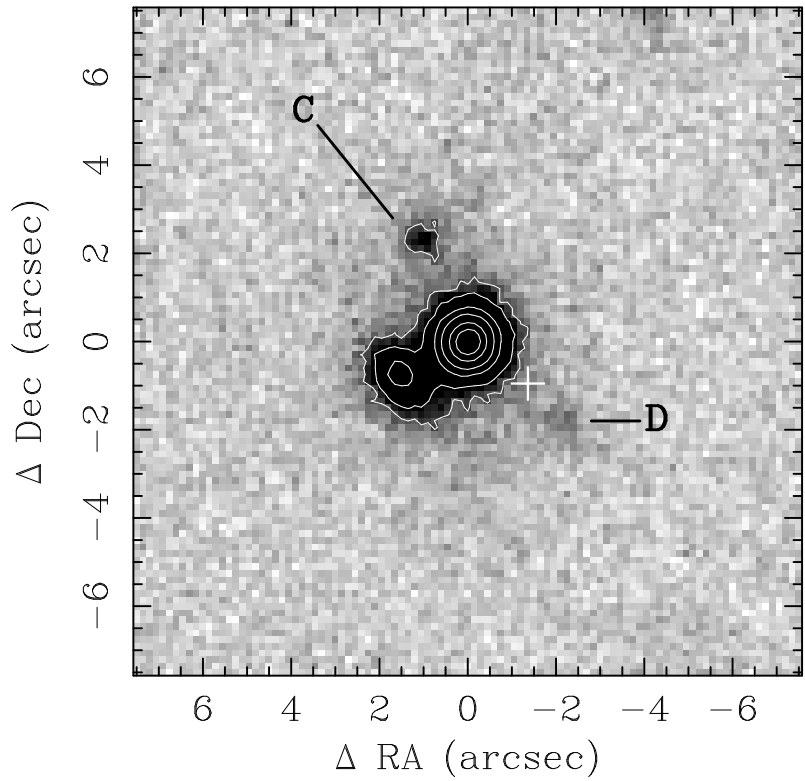
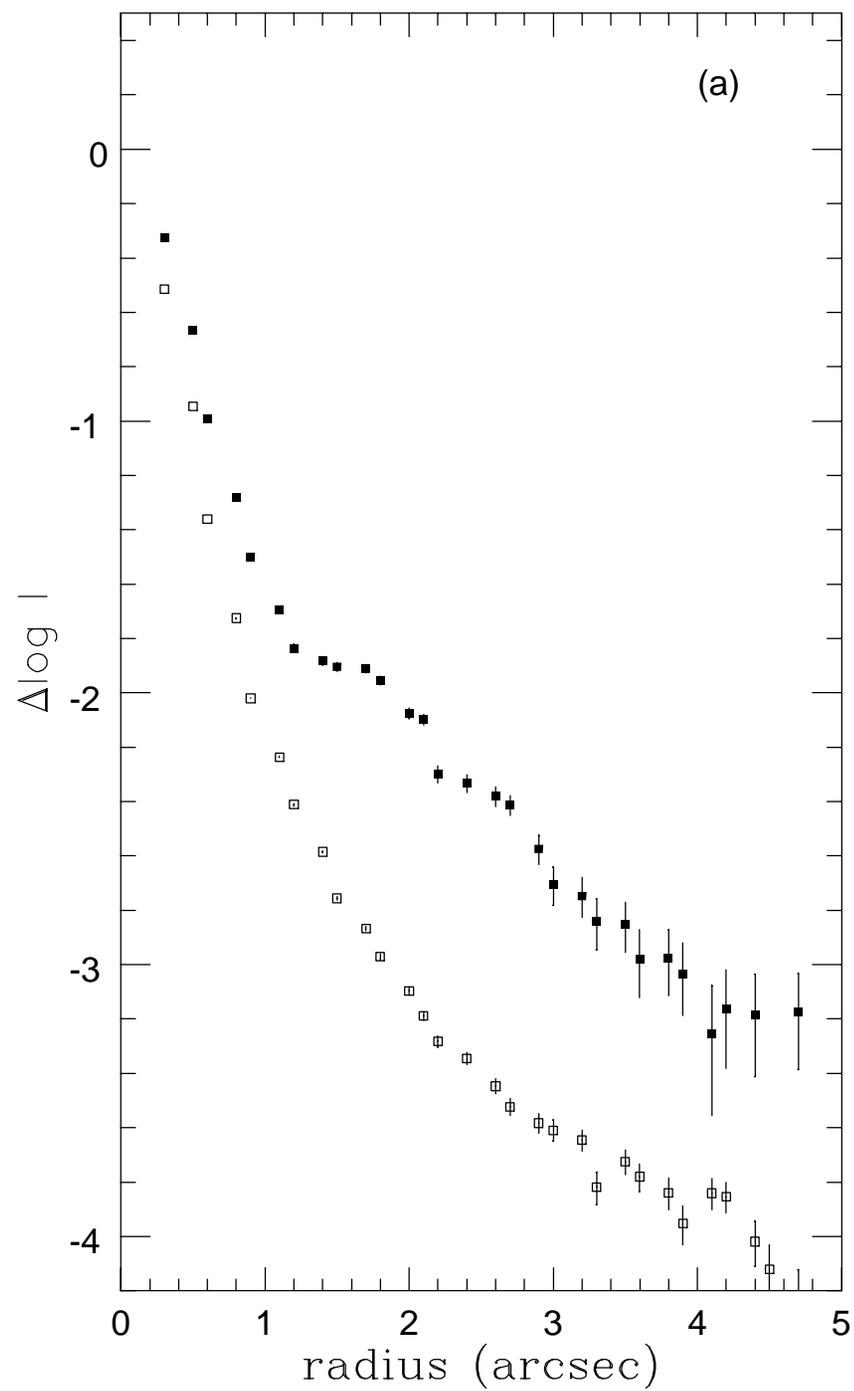
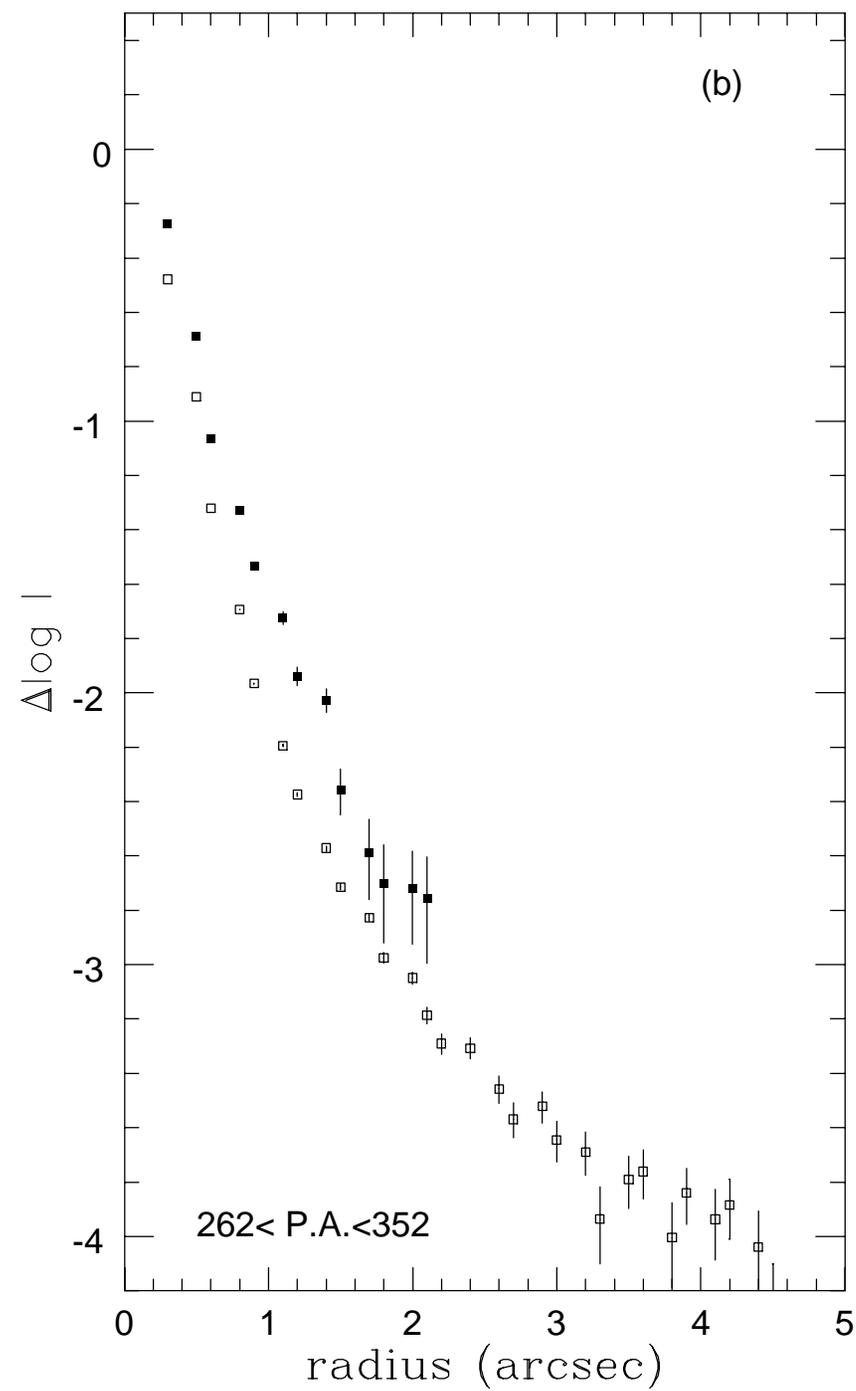
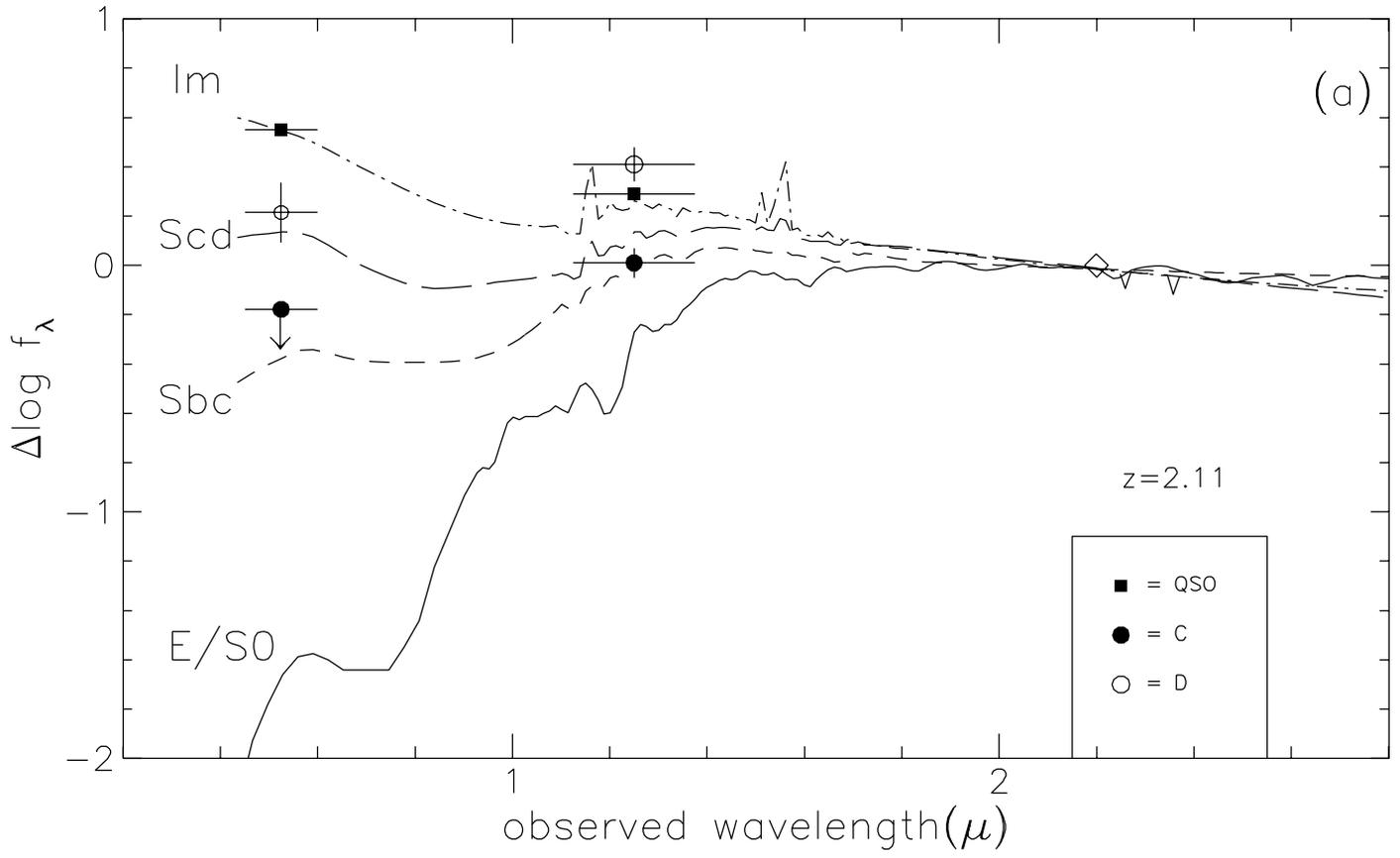
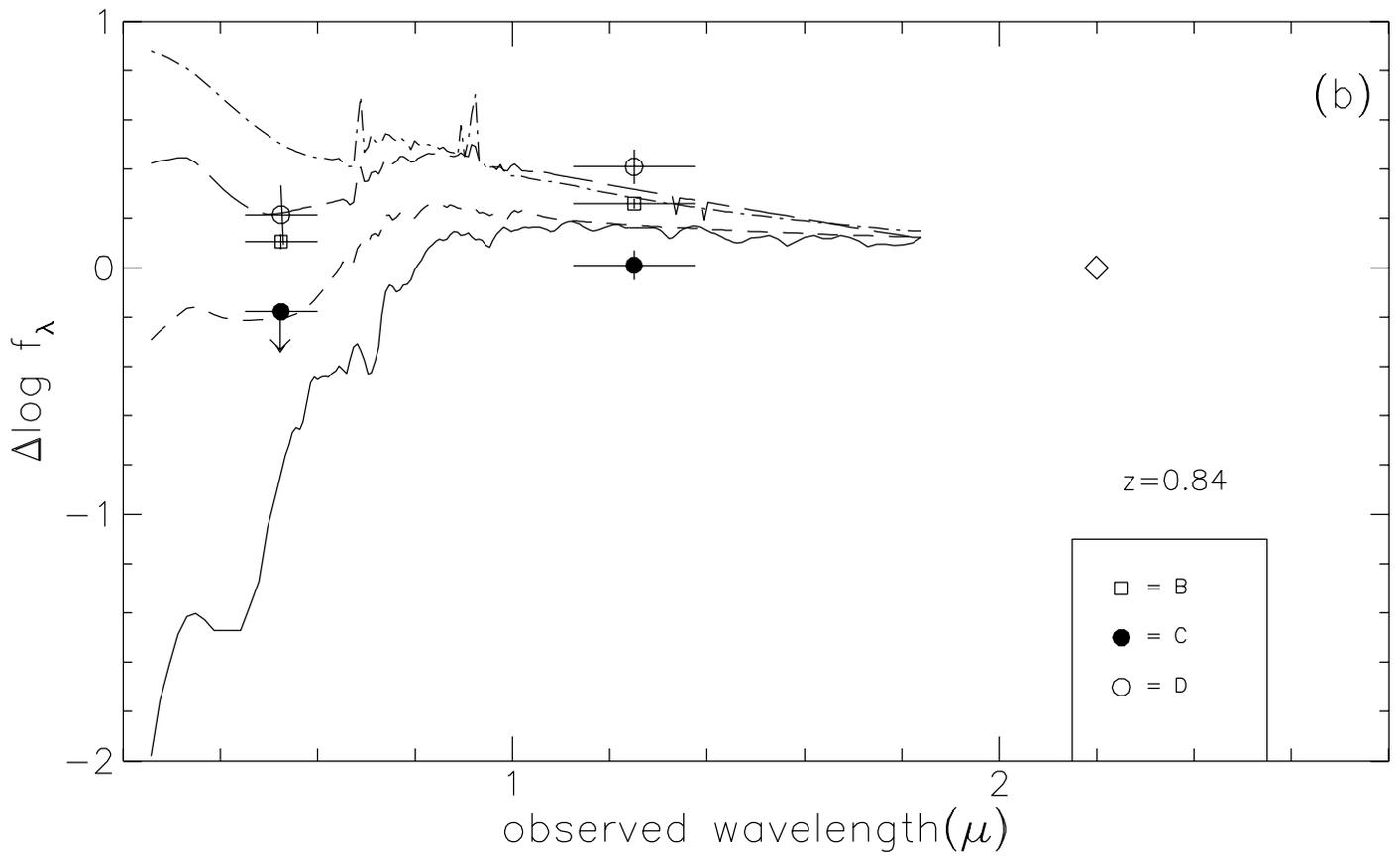
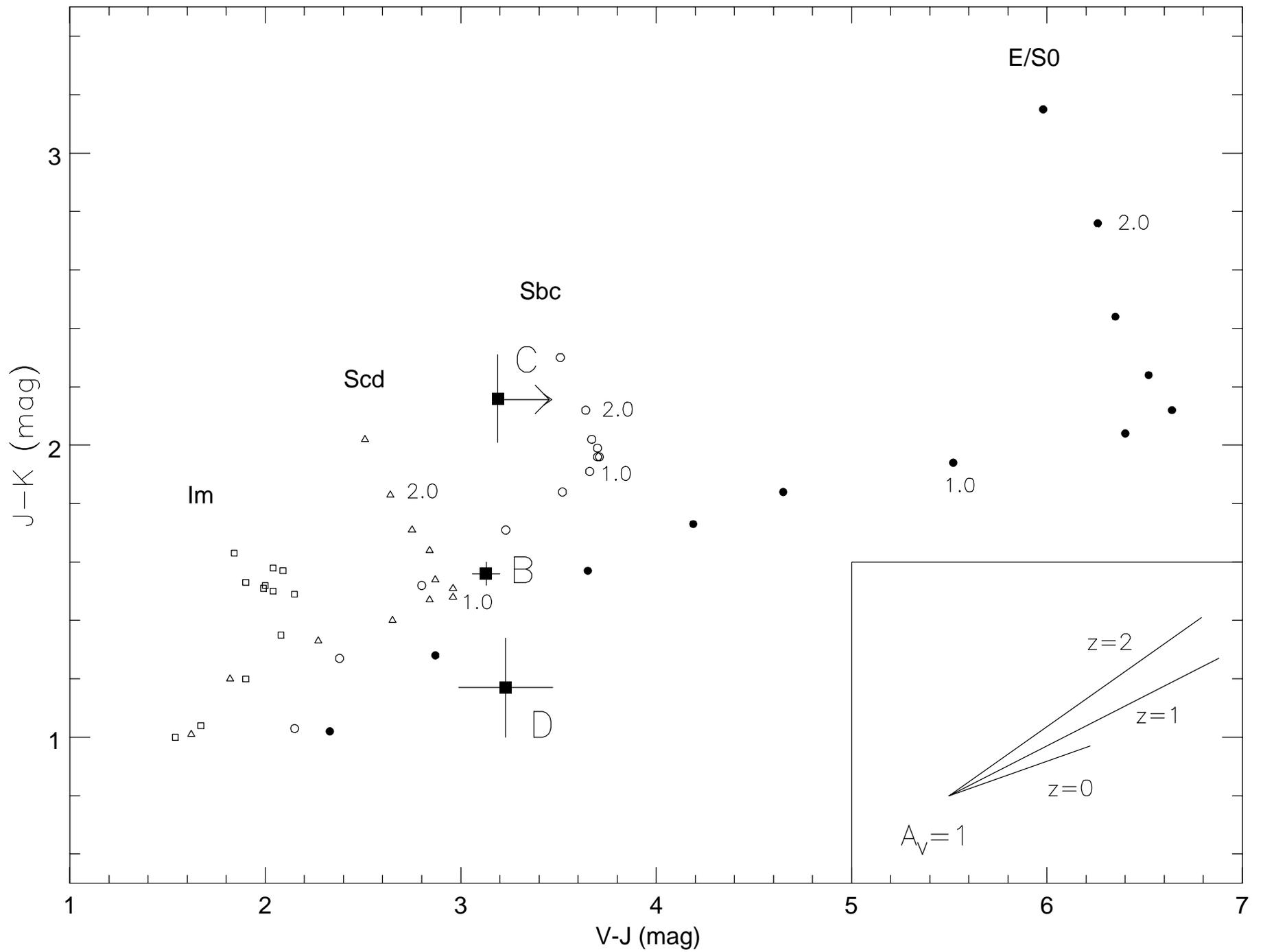
\end{document}